# The origin of the spin glass transition in a model geometrically frustrated magnet

W. Bisson[1,2] & A.S. Wills[1,2,3]

[1]Department of Chemistry, University College London, 20 Gordon Street, London, WC1H 0AJ, UK

[2]Davy-Faraday Research Laboratory, The Royal Institution of Great Britain, 21 Albemarle Street, London, W1S 4BS, UK

[3] The London Centre for Nanotechnology, 17-19 Gordon Street, W1H 0AJ

**Highly frustrated systems have macroscopically degenerate ground states that lead to novel properties. In magnetism its consequences underpin exotic and technologically important effects, such as, high temperature superconductivity[1], colossal magnetoresistence[2], and the anomalous Hall effect[3]. One of the enduring mysteries of frustrated magnetism is why certain experimental systems have a spin glass transition and its exact nature, given that it is not determined by the strength of the dominant magnetic interactions. There have been some suggestions that real systems possess disorder of the magnetic sites or bonds that are responsible[4]. In this letter we show that the spin glass transition in the model *kagomé* antiferromagnet hydronium jarosite arises from a spin anisotropy[5]. This weaker energy scale is much smaller than that of the magnetic exchange, yet it is responsible for the energy barriers that are necessary to stabilise a glassy magnetic phase at finite temperature[5]. The resultant glassy phase is quite unlike those found in conventional disordered spin glasses[4] as it is based on complex collective rearrangements of spins called 'spin folds'[5,6]. This simplifies hugely theoretical treatment of both the complex dynamics characteristic of a spin glass and the microscopic nature of the spin glass transition itself. Understanding the statistical mechanics of this process has implications far beyond the field of magnetism as spin glasses provide important models with which to understand the out-of-equilibrium dynamics that lie at the heart of other frustrated systems, including protein folding[7], neural networks[4] as well as magnets.**

Engineering highly degenerate electronic ground states provides a means to explore the exotic physics associated with out-of-equilibrium dynamics. The simplest examples come from the field of frustrated magnetism where the frustration is generated solely through the geometric construction of the system. This is best exemplified by magnetic spins that are antiferromagnetic coupled to form a triangle, as the geometry prevents a ground state in which all neighbouring moments are related by 180°. Instead a compromise configuration occurs where the spins are mutually oriented at 120°. An important feature arising from this canting is the formation of two ground states that are distinguished by their chirality[8], defined by the vector $\kappa$, taken from the vector products in a clockwise direction:

$$\kappa = \frac{2}{(3\sqrt{3})}[S_1 \times S_2 + S_2 \times S_3 + S_3 \times S_1] \quad (1)$$

The spin configurations that correspond to the two chiralities are indicated by the coloured triangles in Figure 1(a); $\kappa$=+1 (orange), $\kappa$=-1 (green).

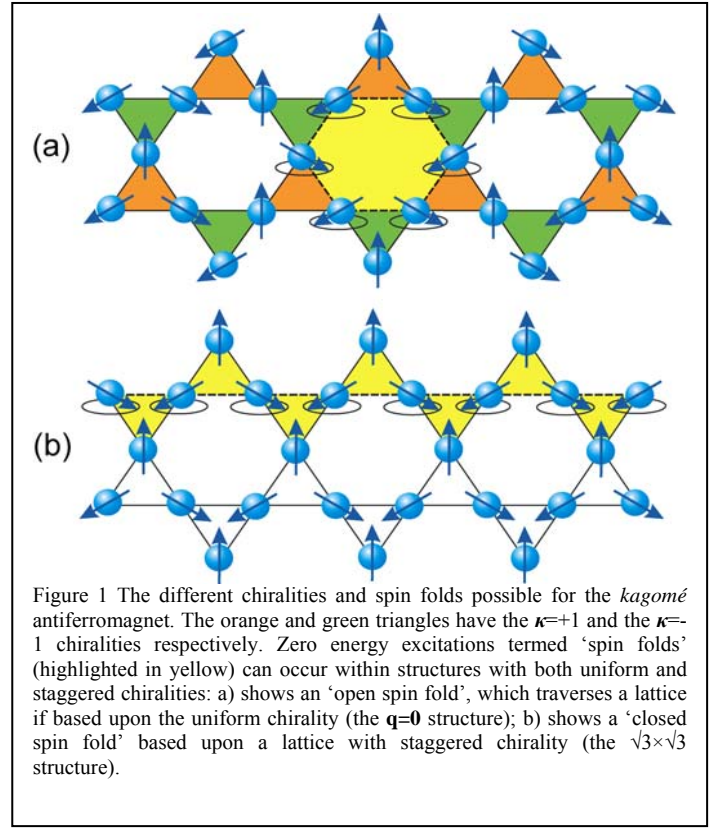

Figure 1 The different chiralities and spin folds possible for the *kagomé* antiferromagnet. The orange and green triangles have the $\kappa$=+1 and the $\kappa$=-1 chiralities respectively. Zero energy excitations termed 'spin folds' (highlighted in yellow) can occur within structures with both uniform and staggered chiralities: a) shows an 'open spin fold', which traverses a lattice if based upon the uniform chirality (the **q=0** structure); b) shows a 'closed spin fold' based upon a lattice with staggered chirality (the √3×√3 structure).

This degeneracy is enhanced by tiling the triangles through vertex sharing to form the 2-D *kagomé* network shown in Figure 1, as the low connectivity of the lattice prevents the rule of a 120° ground state configuration from defining the chiralities of the triangular plaquettes. As each triangle can still be in either $\kappa$ =+1 or -1, this geometrically frustrated antiferromagnet has a degeneracy that scales with the size of this system which can lead to fascinating physics, including spin liquids and spin glasses[9]. The best known example of a *kagomé* antiferromagnet, hydronium jarosite $(H_3O)Fe_3(SO_4)_2(OH)_6$, has attracted considerable attention as it displays an unconventional spin glass-like phase that results from the highly frustrated magnetism of the lattice[10,11]. One of the unsolved problems in frustrated magnetism surrounds the microscopic nature of this glassy magnetic phase and why it is only observed in hydronium jarosite with all the other iron jarosites displaying long range Néel order at low temperature.[12] In this article we show that hydronium jarosite provides a good model for a new type of spin glass, one based on spin anisotropies in a highly degenerate ground state.

Frustration is most simply revealed by the failure of the dominant magnetic interactions to lead directly to magnetic ordering[4]. Instead low symmetry energies, such as further neighbour exchange, are required to break or reduce the degeneracy of the ground state and so lead to magnetic order. The region where the magnetic correlations are strong but order has not yet occurred is called a spin liquid[9] by analogy with structural liquids. Phase transitions from spin liquid to spin glass phases are of considerable interest because such phases are closely related: both possess only short range magnetic order and exhibit highly correlated fluctuations[9]. This relationship is reinforced in the high temperature cuprate superconductors[1,13] where many pass through a spin glass phase prior becoming superconducting and the nature of the superconducting transition itself is similar to theoretical models for 2D quantum spin liquid transitions[9], with magnetic frustration, chiral states and planar (*XY*) anisotropy conspiring together to force the system to behave unexpectedly[14]. Firstly by removing Néel order to form a state of vortices[9] and then with subsequent pairing of two nearby spins to form a singlet to achieve a superconductor ground state proper[1]. The degeneracy and chirality of the *kagomé* antiferromagnet makes it an ideal model system with which to study the roles of energy scales, such as anisotropy, in exotic fluctuating states.

An important aspect of the degenerate manifold of the *kagomé* antiferromagnet is that it is connected, that is to say, any ground state can be converted into any other through a series of collective spin reorientations that never break the 120° rule for a ground state. These zero energy modes are termed 'spin folds'[5] and are lines of two sublattices that rotate according to the axis defined by the third sublattice. Spin folds can either form closed loops[5] in regions of staggered chirality or extended lines in regions of uniform chirality (Figure 1) and their formation flips the chiralities of the involved triangles, allowing the system to explore all ground states in the degenerate manifold.[11] In the case of isotropic spins this fluid motion of spin reorientations persists on cooling, causing the Heisenberg *kagomé* antiferromagnet to remain a spin liquid even down to *T*=0. Planar, *XY,* anisotropy, however, creates an energy cost to any spin reorientation out of the *kagomé* plane that retards the fluid motion of spin folds, giving rise to the complex out-of-equilibrium dynamics that are characteristic of spin glasses[5]. The spin glass transition then corresponds to a type of Kosterlitz-Thouless (KT) transition where intersecting spin folds create vortex-like point defects that bind together in the low temperature phase.[5,8] Even a weak anisotropy has a profound effect in the *kagomé* antiferromagnet as it produces a rapid crossover from a low density of defects to a high density of bound defects and consequently to a critical transition[5].

The nature of the *kagomé* spin glass phase has several contrasts with more common site-disordered spin glasses as it is based on an evolution through a degenerate manifold. Firstly, the successive creation of the spin folds is non-Abelian, meaning the spin configuration formed by successive folds labelled *A* and then *B* is not the same as that resulting from *B* and then *A*– a spin algebra that causes the system to have a memory of all the ground states it has traversed[5,11]. Secondly, the spin configurations involved retain the 120° ground state rule all times. This means that the energy landscape of the magnet is translationally uniform, despite the disorder in the spin chirality- a remarkable symmetry that allows a hidden channel for the low energy Goldstone modes familiar in ordered magnets to occur in the *kagomé* spin glass, though in this case they represent excitations from an ordered energy landscape rather than a translationally ordered spin structure.[18]

Hydronium jarosite has attracted considerable attention because of the stark contrasts between its glassy spin dynamics and thermodynamics, with those of more common site-disordered spin glasses[5,10,11]. Two of its most remarkable properties are its magnetic specific heat that is quadratic with temperature, characteristic of Goldstone modes in a 2-D antiferromagnet[5,10], rather than the linear response expected for a disordered spin glass[4], and the extensive memory of spin relaxations in temperature cycling measurements that would be expected for non-Albelian dynamics[11]. The elegant ability of this simple model to explain, what were at the time, extremely surprising properties promotes hydronium jarosite as an important model system with which to explore both the nature of spin glasses and the way in which macroscopic degeneracies can lead to complex out-of-equilibrium phases.

To confirm whether the spin glass state in hydronium jarosite is the result of an anisotropy energy we characterised the relationship between local coordination of the magnetic $Fe^{3+}$ ions determined by single crystal X-ray diffraction and the critical spin glass transition, $T_g$, of a number of hydronium jarosite prepared by hydrothermal methods[16].

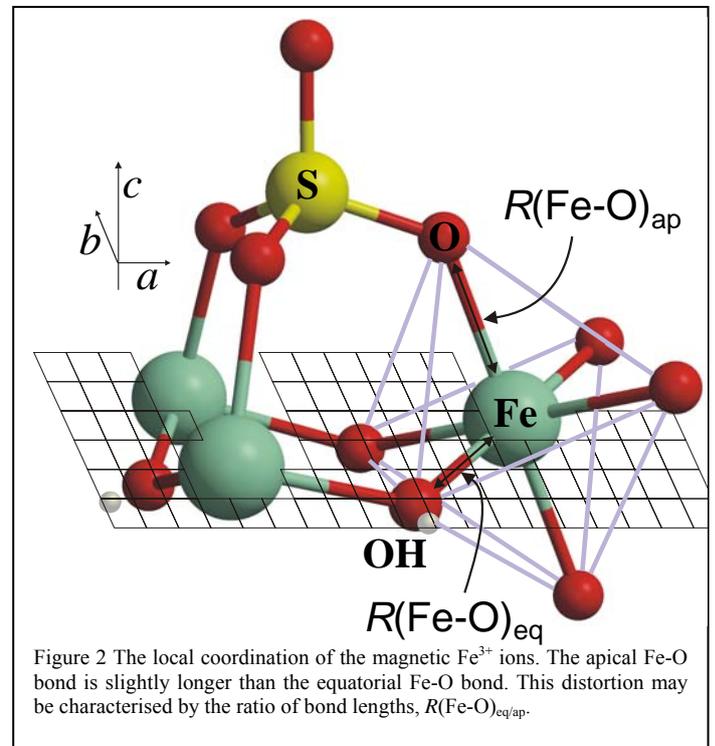

Figure 2 The local coordination of the magnetic $Fe^{3+}$ ions. The apical Fe-O bond is slightly longer than the equatorial Fe-O bond. This distortion may be characterised by the ratio of bond lengths, $R$(Fe-O)$_{eq/ap}$.

The jarosite crystal structure is best described in the space group $R\bar{3}m$ and has the general formula $AFe_3(SO_4)_2(OH)_6$, (A = $H_3O^+$, $NH_4^+$, $K^+$, $Na^+$, $Rb^+$, $Ag^+$, $½Pb^{2+}$ or $½Tl^{2+}$).[17] The $Fe^{3+}$, $S=5/2$, ions make up a series of translationally related *kagomé* layers with an …ABC… stacking arrangement. These layers are sufficiently well separated, $R$(Fe-Fe)$_{interlayer}$~5.64Å, that the magnetic Hamiltonian may be considered essentially 2D, which is confirmed by our following observations. The magnetic exchange between Fe ions is mediated through the bridging hydroxyl groups (Figure 2) and the nearest neighbour distance is $R$(Fe-Fe)$_{intralayer}$~3.67 Å. There is a markedly different behaviour between the hydronium and non-hydronium jarosites with all of the non-hydronium jarosites ordering at low temperatures into a long-ranged Néel state with the propagation vector ***k***=(0 0 3/2), with respect to the hexagonal setting of the space group.[12] Whether or not this ordering process is one or two staged appears to depend on the samples. In all cases, however, the lower temperature transition occurs at $T_{N2}$ ~55K[12].

Hydronium jarosite, is quite different from the non-hydronium jarosites as it displays a critical spin glass transition at $T_g$~17K. In this work our synthesis methods produced samples with a range of spin glass freezing transitions, $T_g$, between 11 and 20K. The single crystals formed were large enough for single crystal X-ray diffraction studies were collected at 80K using Mo K$\alpha_1$ radiation (0.71073 Å). While the small changes in stoichiometry of these samples were too small to be resolved by chemical analysis or crystal structure refinement, the latter confirmed that the Fe occupancy was >94% for all samples studied.

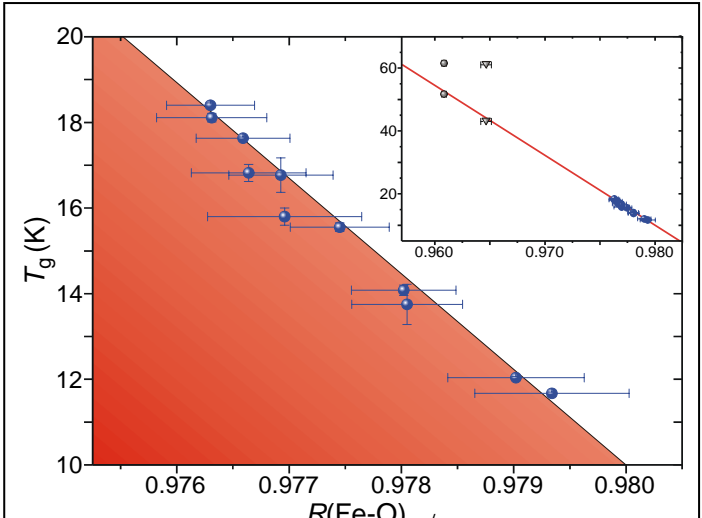

Figure 3. The relation between the critical spin glass freezing transition in hydronium jarosite, $T_g$, and the ratio of bond lengths, $R$(Fe-O)$_{eq/ap}$, the more symmetrical the Fe octahedra, the lower the transition temperature. he approximately linear correlation suggests that the anisotropy is proportional to the extent of the distortion. The inset shows that the same linear function fitted for hydronium jarosite can be extended to the antiferromagnetic Néel ordering temperatures of KFe$_3$(SO$_4$)$_2$(OH)$_6$ (grey circles) and (NH$_4$)Fe$_3$(SO$_4$)$_2$(OH)$_6$ (grey triangles).

Detailed studies of the single crystal diffraction data reveal small changes that arise from a strong correlation between the geometry of the Fe coordination octahedra and the value of $T_g$. Looking in detail at the Fe$^{3+}$ coordination, the Fe$^{3+}$ ions have point symmetry $2/m$ as then reside on the $9d$ site of the $R\bar{3}m$ space group and are coordinated by 4 hydroxyl oxygen atoms in the equatorial (eq) plane and 2 apical (ap) oxygen atoms from the sulphate groups as shown in Figure 2. The apical Fe-O bond is slightly longer than the equatorial bond and the distortion away from octahedral symmetry may be characterised by the ratio of bond lengths, $R$(Fe–O)$_{eq/ap}$=$R$(Fe–O)$_{eq}$/$R$(Fe–O)$_{ax}$. Figure 3 shows an apparently linear dependency of $T_g$ on the ratio $R$(Fe-O)$_{eq/ap}$, with the samples that have more distorted Fe-octahedra displaying higher values of $T_g$. It should be noted that the expansion along the crystallographic $c$-axis that is concomitant with the increasing distortion further pushes apart the *kagomé* layers, thereby reducing the strength of the interlayer coupling. This property both indicates that interlayer interactions are not responsible for the spin glass phase and stresses the contrasts with conventional disordered spin glass systems, as these do not possess critical phase transitions at finite temperature.

This relationship between the local distortion of the Fe$^{3+}$ and the spin glass transition is reinforced by studies of the non-hydronium jarosites which show them all to possess far greater distortions away from octahedral symmetry and correspondingly higher antiferromagnetic Néel transition temperatures. The inset of Figure 3 shows an extrapolation of the linear dependence of $T_g$ with $R$(Fe–O)$_{eq/ap}$ observed in hydronium jarosite. Excellent agreement between this trend and the lower temperature transitions of KFe$_3$(SO$_4$)$_2$(OH)$_6$ ($T_{N1}$=51.7 (5), $T_{N2}$=61.5 (5), $R$(Fe-O)$_{eq/ap}$=0.96084 (20), grey circles) and (NH$_4$)Fe$_3$(SO$_4$)$_2$(OH)$_6$ ($T_{N1}$=43.0 (5), $T_{N2}$=61.0 (5), $R$(Fe-O)$_{eq/ap}$ =0.96466 (48), grey triangles) suggests that the orderings in the hydronium and the non-hydronium iron jarosites share a common energy scale which is quite remarkable given the different natures of the orderings.

In light of the observation of anisotropy in the *kagomé* spin glass hydronium jarosite, we will now consider whether its origin as it could be due to either crystal field (CF) anisotropy or the Dzyaloshinsky-Moriya Interaction (DMI)[18,19]. As Fe$^{3+}$ would naïvely expected to be $S=5/2$, $L$=0 both of these interactions should be zero as they require spin-orbit coupling to be present. This is not the case as EPR and Mössbauer spectroscopy measurements have indeed shown that the Fe$^{3+}$ spins in hydronium jarosite lie in the *kagomé* plane[20] evidencing a mixing in of some orbitally degenerate excited electronic terms into that of the Fe$^{3+}$ Such mixing, in addition to spin-orbit coupling, has also deduced from hyperfine splitting in Mössbauer spectroscopy and ground state splitting in EPR for Fe$^{3+}$ in yttrium iron garnet [21,22]. In the jarosites, these contributions will be enhanced by increasing large distortion away from perfect octahedral symmetry, compatible with Figure 3.

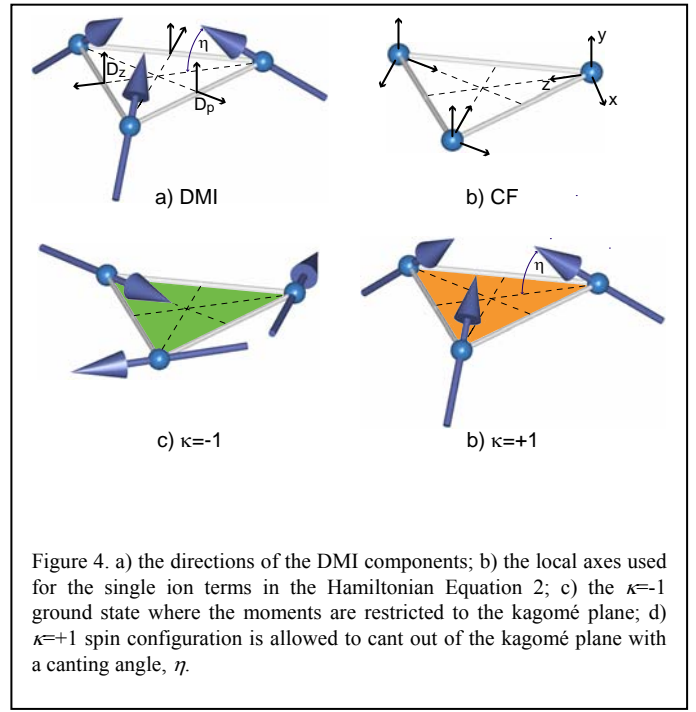

Figure 4. a) the directions of the DMI components; b) the local axes used for the single ion terms in the Hamiltonian Equation 2; c) the $\kappa$=-1 ground state where the moments are restricted to the kagomé plane; d) $\kappa$=+1 spin configuration is allowed to cant out of the kagomé plane with a canting angle, $\eta$.

Inspection reveals that planar anisotropies will have quite different effects on the *kagomé* spin glass depending on whether they are due to CF effects or the DMI. This is because evolution of this glassy phase occurs through the creation of spin folds that flip the chiralities of the triangles that are involved. Consequently, for an anisotropy to stabilise the spin glass phase, it must be degenerate with respect to both $\kappa$=±1 plaquettes. If this were not the case, the system would simply crystallise with order corresponding to the lowest energy chiral plaquettes. If we consider the generic Hamiltonian[23]

$$H = \sum_{nn}\left[J_1 \mathbf{S}_i \cdot \mathbf{S}_j + \mathbf{D}_{ij} \cdot \mathbf{S}_i \times \mathbf{S}_j\right] + \sum_{nnn} J_2 \mathbf{S}_k \cdot \mathbf{S}_l + D\sum_i \left(S_i^z\right)^2 - E\sum_i \left[\left(S_i^y\right)^2 - \left(S_i^{x'}\right)^2\right] \quad (2)$$

where $\mathbf{D}_{ij}$ is the DMI vector which acts in a plane perpendicular to the $ij$ bond within the jarosite structure shown in Figure 4 and the single-ion anisotropy terms are $D$ and $E$ defined with respect to local axes: $x'$ and $y'$ in the kagomé plane, and $z'$ being perpendicular to it. The single-ion terms are second order relative to DMI as denoted by the next nearest neighbour (nnn) term. $\mathbf{D}_{ij}$ may in turn be separated into two components: $D_p$ which lies in the $xy$ plane kagomé and $D_z$ the perpendicular component. Analytical and Monte Carlo calculations have shown that the chirality of a triangle depends on the magnitude and direction of the components of $\mathbf{D}_{ij}$[24] with the sign of $D_z/J_1$ both dictating the chirality and acting as an XY anisotropy that forces the moments into the kagomé plane. A positive $D_z/J_1$ with a non-zero $|D_p|$ will select the $\kappa$=+1 chirality leading to the $q=0$ spin structure shown in Figures 1b and 4 with the moments allowed to cant out of the plane: there is no canting when $|D_p|$=0. While a small negative $D_z/J_1$ can support a $\mathbf{q=0}$ structure with canting if $|D_p/J_1|$ is similar or larger in magnitude, a large negative $D_z/J_1$ will result in $\mathbf{q=0}$ coplanar structure based upon the $\kappa$=-1 chirality regardless of $|D_p/J_1|$. Therefore a spin glass phase based on both $\kappa$=+1 and -1 chiralities in which the moments lie in the kagomé plane requires that $D_z/J_1$=0 and that the inplane component of $\mathbf{D}_{ij}$ be zero, $|D_p|$=0. Consequently, the presence of DMI in the jarosites is found to be incompatible with the kagomé spin glass phase. It is however compatible with CF terms if $D$>0 and $E$=0 (or $E$ is small with respect to fluctuations) as these support the degeneracy of the planar $\kappa$=±1 configurations.

Unfortunately, the nature of a spin glass prevents the determination of $D$ and $E$ from spin wave measurements, so information about them must be inferred from comparison with the other iron jarosites that show long-range order. Recent inelastic neutron scattering from single crystals of KFe$_3$(SO$_4$)$_2$(OD)$_6$ have suggested that it is the DMI terms lead to long range order in this material.[23,26] Pivotal to the analysis by Matam *et al* is the assumption that the moments are canted to form an umbrella structure at low temperature, based on an extrapolation of magnetisation data taken at $T$=50K,[25] as within the DMI model this reproduced properties of the spin wave data such as the splitting at the $\Gamma$ point.[26] The more in depth treatment by Yildirim and Harris[23] shows that the value of this canting angle is important in the spin wave analysis and that if the moments are allowed to lie in the kagomé plane during the fitting procedure, the CF model can indeed fit the data equally well as the DMI model. Supporting this picture of the moments being in the plane is the observation of a spin flop transition at $T_{N1}$ where the moments flop towards the kagomé plane[27]. Exploration of the CF terms in both the spin glass and Néel states indicates that the uniaxial $E$ term will break the degeneracy of the $\kappa$=±1 configurations just as the DMI does and that the observed differences between hydronium and non-hydronium jarosites can be understood simply if the smaller distortion in hydronium jarosite is associated with a smaller $E$ term, than those in more distorted jarosites which display Néel order. The crystal structure of hydronium jarosite is then fortuitous in that it possesses the planar anisotropy $D$ term that is accompanied by a small $E$ as this situation supports the formation of the kagomé spin glass phase. In the other iron jarosites the larger value of $E$ prevents the formation of the spin glass phase by stabilising conventional Néel order.

In summary, we show that the kagomé antiferromagnet hydronium jarosite displays a glass-like magnetic state as the result of a weak planar single-ion anisotropy which retards the evolution through the highly degenerate ground state. As the local coordination of the Fe$^{3+}$ ion becomes more distorted, enhancement of the single ion terms stabilises the spin glass state at higher temperatures until the axial $E$ term induces long-range Néel order. The stark simplicity of this model kagomé antiferromagnet is in contrast with conventional disordered spin glasses. It may therefore provide important insights into one of the long standing questions in condensed matter physics- the nature of spin glass transition- and help our understanding of the analogous out-of-equilibrium dynamics found in diverse systems such as spin glasses, high $T_c$ superconductors, neural networks and protein folding.

**Method section**

The hydronium jarosites were synthesised under standard hydrothermal synthesis conditions using Pyrex tubes with PTFE screw caps. 2g of Fe$_2$(SO$_4$)$_3$ were dissolved in 15 ml of water/methanol solutions and heated between 120 and 150°C for 21 hours. The non-hydronium jarosites were synthesised using redox methods where iron wire is oxidized. The following K$_2$(SO$_4$)$_2$ (2.44g, 0.28mol), and (NH$_4$)$_2$(SO$_4$)$_2$ (1.85g, 0.28mol) were each dissolved and made up to 25cm$^3$ with distilled water, to which 1.1cm$^3$ of concentrated H$_2$SO$_4$ was added. For each reaction 0.336g of iron wire, 2mm diameter, 99.9% were put with the relevant A-site sulfate solution into a pressure tube (38cm$^3$ total capacity). The reaction took place at 170°C over 48 hours. The magnetic susceptibility data were collected using a Quantum Design SQUID magnetometer in a measuring field of 100 G. Single crystal X-ray diffraction data were taken at 80(2)~K using an Oxford Cryostream. The data for the potassium jarosite was taken from SRS facility collected at Daresbury with X-rays of wavelength 0.6893 Å at 85(2)K Position and peak intensities were extracted from the raw data using DENZO SMN and scaled using SCALEPACK; SADABS was used for adsorption correction. SHELX-97[28] was used for structure solution and refinement.

**Acknowledgements**
This work has been funded by the Royal Society and EPSRC grant number EP/C 534654. We would like to thank P.C.W. Holdsworth and S.T. Bramwell for discussions and the ENS Lyon for provision of travel funds. The UK National Crystallography Service (NCS) and the assistance of. S. Coles is gratefully acknowledged.